\documentclass{book}
\usepackage[contrib,lang=british]{ems-book} %% change to `american' if you use American English

\begin{document}
\mainmatter

%------
% Insert the title of your paper and (if necessary)
% a short title for the running head.
%------
\title{Quantum number towers for the Hubbard and Holstein models}
\titlemark{Quantum number towers for the Hubbard and Holstein models}

%------
% Insert full names of the authors.
% Add further authors as follows:
%  \emsauthor{2}{}{}
%  \emsauthor{3}{}{}
% etc.
% Abbreviate first names for the running head.
%------
\emsauthor{1}{James K. Freericks}{J.~K.~Freericks}

%------
% Use \authormark if the list of authors is too
% long for the running head: \authormark{A.~Doe et al.}
%------

%------
% Add one \emsaffil and one \email for each author.
% NOTE: The address does NOT appear in the paper.
% It will probably be printed in an appendix.
%------
\emsaffil{1}{Department of Physics, Georgetown University, 37th and O Sts. NW, Washington, DC 20057, USA \email{james.freericks@georgetown.edu}}

%------
% Add MSC 2020 codes according to www.ams.org/msc/msc2020.html.
% Secondary codes (in square brackets) are optional.
%------

%------
% Add a list of keywords.
%------

%------
% Insert your abstract.
%------
\begin{abstract}
In 1989, Elliott Lieb published a \textit{Physical Review Letter} proving two theorems about the Hubbard model. This paper used the concept of spin-reflection positivity to prove that the ground state of the attractive Hubbard model was always a nondegenerate spin singlet and to also prove that the ground state for the repulsive model on a bipartite lattice had spin $\big||\Lambda_A|-|\Lambda_B|\big|/2$, corresponding to the difference in number of lattice sites for the two sublattices. In addition, this work relates to quantum number towers---where the minimal energy state with a given quantum number, such as spin, or pseudospin, is ordered, according to the spin or pseudospin values. It was followed up in 1995 by a second paper that extended some of these results to the Holstein model (and more general electron-phonon models). These works prove results about the quantum numbers of these many-body models in condensed matter physics and have been very influential. In this chapter, I will discuss the context for these proofs, what they mean, and the remaining open questions related to the original work. In addition, I will briefly discuss some of the additional work that this methodology inspired.
\end{abstract}

\makecontribtitle

%------
% INSERT THE BODY OF THE PAPER HERE (except
% acknowledgments, funding info and bibliography)
%------
\section{Introduction}

In the late 1950s and early 1960s two models for condensed matter physics systems with electron-phonon coupling~\cite{holstein} (called the Holstein model) and electron-electron interaction~\cite{hubbard} (called the Hubbard model) were proposed. These models were simple in structure and easy to describe. But, they both turned out to be extremely difficult to solve. 

In the late 1960's Gaudin~\cite{gaudin} and Yang~\cite{yang} solved a long-standing problem in determining how to extend the Bethe-ansatz~\cite{bethe} from spin models to fermions. Shortly thereafter, Lieb and Wu~\cite{lieb-wu} solved the one-dimensional Hubbard model problem and showed that the Mott metal-insulator transition occurred for infinitesimally large $U$ at half filling by showing a gap opened in the density of states  for any positive value of the interaction $U$. But solutions outside of one dimension proved difficult to find. In the early 1990s, dynamical mean-field theory was introduced as a way to solve strongly correlated models in infinite spatial dimensions. The properties of the Hubbard model were mapped out over a decade or so using increasingly more sophisticated numerical methods~\cite{kotliar-review}. Results for the Holstein model are even meagerer. There are no known exact solutions, although numerical work in one-dimension~\cite{holstein-dmrg} and in dynamical mean-field theory~\cite{freericks} have established some properties of the model.

Elliott Lieb's work on these two models is important, because it showed that one can use rigorous analytical methods to understand properties of the quantum numbers of the system---constraints on the quantum numbers of the ground-state, and ``tower'' structures in the energy eigenvalue spectrum, similar to the Lieb-Mattis proof of the absence of ferromagnetism in one-dimensional models~\cite{lieb-mattis}. This work was based on the concept of \textit{spin-reflection positivity}, a powerful new idea in many-body physics that allows one to show precisely when electrons of opposite spins prefer to be located on the same lattice site due to an attractive interaction. Then a partial particle-hole transformation allows one to establish consequences of these results for the repulsive model. The Hubbard model work was completed in 1989~\cite{lieb-hubbard} and was refined and extended to electron-phonon coupled models in 1995~\cite{freericks-lieb}.

The Hubbard and Holstein models are models of electron correlations on lattices (technically speaking the lattices need not be periodic, and can be thought of just as a collection of sites $\{x\mid x\in \Lambda\}$, with ``bonds'' corresponding to pairs $x\in \Lambda$ and $y\in \Lambda$ where a hopping matrix $t_{xy}\ne 0$); we let $|\Lambda|$ denote the (finite) number of lattice sites in $\Lambda$. Then the Hubbard model involves electrons that hop on the ``lattice'' and interact when two electrons of opposite spin sit on the same lattice site. It is defined as
\begin{equation}
\hat{\mathcal{H}}=\sum_\sigma\sum_{x,y\in\Lambda}t_{xy}\hat{c}_{x\sigma}^\dagger\hat{c}_{y\sigma}^{\phantom{\dagger}}+\sum_{x\in\Lambda}U_x\hat{n}_{x\uparrow}\hat{n}_{x\downarrow}.
\label{eq:hubbard}
\end{equation}
Here, we use fermionic creation (annihilation) operators $\hat{c}_{x\sigma}^\dagger$ ($\hat{c}_{x\sigma}^{\phantom{\dagger}}$) for a fermion of spin $\sigma$ at site $x$. The ``lattice'' is a collection of sites, which does not need to have any structure to it. The hopping matrix is arbitrary except it is real-valued and $t_{xy}=t_{yx}$; it is not allowed to depend on spin, but it is allowed to have diagonal elements, in that one can have $t_{xx}\ne 0$. The fermionic operators satisfy the ordinary anticommutation relations given by 
\begin{equation}
    \left \{\hat{c}_{x\sigma\phantom{'}}^{\phantom{\dagger}},\hat{c}_{y\sigma'}^\dagger\right\}=\delta_{xy}\delta_{\sigma\sigma'}~~\text{and}~~\left\{\hat{c}_{x\sigma\phantom{'}}^{\phantom{\dagger}},\hat{c}_{y\sigma'}^{\phantom{\dagger}}\right\}=0.
\end{equation}
The number operators are $\hat{n}_{x\sigma}=\hat{c}_{x\sigma}^\dagger\hat{c}_{x\sigma}^{\phantom{\dagger}}$. While the above form is quite general, in condensed matter physics, the Hubbard model is most often studied on a periodic lattice with nearest-neighbor hopping (sometimes longer range too) and with an interaction $U$ that does not depend on the lattice site.

Note that this model is a significant simplification of the model of a real material. This is because we consider only one band and we only have an on-site Coulomb interaction. It is well-known that the Coulomb interaction is fairly long-range, so it often involves interactions farther away than just on-site---the idea of Hubbard was that screening could make the Coulomb interaction most important only when both electrons are in the same unit cell. The other thing it misses is the exchange interaction. This can be an important interaction, but it requires multiple bands at a given lattice site to have exchange interaction effects---as a result, this model tends to de-emphasize ferromagnetism, which is often enhanced by the degeneracy associated with multiple bands; but in special cases the Hubbard model does have ferromagnetic solutions.

The other model we will consider is a general electron phonon coupling model (which includes the Holstein model, but much more), given by electrons that hop on a lattice and interact with phonon modes through a coupling of the electron charge at a lattice site to a function of the phonon coordinates and via a modulation of the hopping. The Hamiltonian is given by
\begin{align}
    \hat{\mathcal{H}}_{ep}&=\sum_\sigma\sum_{xy\in\Lambda}t_{xy}(\hat{\vec{q}})\hat{c}_{x\sigma}^\dagger\hat{c}_{y\sigma}^{\phantom{\dagger}}+\sum_{x\in\Lambda}G_x(\hat{\vec{q}})(n_{x\uparrow}+n_{x\downarrow})+\sum_{i=1}^\nu\left (\frac{\hat{p}_i^2}{2m_i}+\frac{1}{2}m_i\omega_i^2\hat{q}_i^2\right )\nonumber\\
    &+V_{an}(\hat{\vec{q}}).
    \label{eq:el-ph}
\end{align}
There are many more conditions to discuss about this Hamiltonian. The phonon operators include position $\hat{q}_1,\cdots,\hat{q}_\nu$ and momentum $\hat{p}_1,\cdots,\hat{p}_\nu$, which satisfy the canonical commutation relation $[\hat{q}_j,\hat{p}_{k}]=i\delta_{jk}$ (we set $\hbar=1$). We collectively refer to these operators as $\hat{\vec{q}}$ and $\hat{\vec{p}}$. The number of phonon modes is not assumed to be linked to the lattice sites, although for the Holstein model, it is, with one phonon mode per lattice site. We will work in the coordinate representation, where $\hat{q}_j\to q_j$ and $\hat{p}_j\to -i\frac{\partial}{\partial q_j}$ and these operators act on functions in $L^2(\mathbb{R}^\nu)$.  $V_{an}(\vec{q})$ is the anharmonic part of the phonon potential (all nonquadratic terms); it is bounded from below $V_{an}(\vec{q})\ge C$, and goes to infinity faster than linearly.

The hopping matrix depends on $\vec{q}$ (but not on momentum) and for any fixed $\vec{q}$, the hopping matrix is real and symmetric $t_{xy}(\vec{q})=t_{yx}(\vec{q})$ and the trace satisfies $\text{Tr} |t_{xy}(\vec{q})|<\infty$. The functional dependence is that $t_{xy}(\vec{q})$ is an arbitrary measurable real-valued function of $\vec{q}$. Similarly $G_x(\vec{q})$ is an arbitrary real-valued function of the coordinates $\vec{q}$. We also will require a boundedness from below of the total phonon energy given by
\begin{equation}
    -2\text{Tr}|t(\vec{q})|-2\sum_{x\in\Lambda}|G_x(\vec{q})|+\frac{1}{2}\sum_{i=1}^\nu m_i\omega_i^2+V_{an}(\vec{q}).
    \label{eq:bounded}
\end{equation}

The general electron-phonon problem includes both the Su Schreiffer Heeger model \cite{ssh} and the Holstein model~\cite{holstein} (and even more general models). For the former, one needs to express the system in a normal mode basis (which is not how the model is usually written down) with no anharmonic potential, while for the latter, we associate a constant frequency harmonic phonon with each lattice site.

The Holstein model is also a simplification---it has no acoustic phonons in it---only Einstein modes. Hence, it is not as accurate in describing heat transport. In the generalization we describe above, we have much more generality, but even here, it is not the most general case one can consider, as it ignores nonlinear electron-phonon coupling and some additional anharmonic effects.

These models have a rich history of behaviors too voluminous to completely discuss here, but we will describe some of the general behavior that has been seen via different numerical methods (much of this behavior cannot be proven to occur in general). In the Hubbard model, the repulsive case has a Mott metal-insulator transition at half filling, and an antiferromagnetic spin-density wave phase at low temperature. Away from half filling, a d-wave superconductor is expected to occur in two dimensions. For the attractive case, the system has a superconducting ground state, which becomes degenerate with a charge-density-wave phase at half filling. The Su-Schrieffer-Heeger model and the Holstein model both have charge-density-wave behavior at half filling and are superconducting away from half filling. They all can possess more complex orders as well, especially in two dimensions.

The proofs that Elliott Lieb worked out relate to the ground-state quantum numbers for finite-sized systems. The proofs work with much more general versions of the models than what are usually studied in condensed matter physics. The main element in these proofs relies on a new concept called spin-reflection positivity, and it is used to show that the attractive interaction between electrons favors having two electrons sitting on the same lattice site. The application to repulsive models arises from partial particle-hole transformations and hence they apply only to the Hubbard model at half filling. Nevertheless, these proofs have led to tremendous insight about these models and have opened up new areas of research that remain active to this day. We will discuss more about that later in the paper.

But first, we must discuss the two sets of operators that commute with these Hamiltonians, for the proofs are all about the quantum numbers of these operators. To begin, both models conserve the total number operator for each spin $\hat{N}_\sigma=\sum_{x\in\Lambda}\hat{c}_{x\sigma}^\dagger\hat{c}_{x\sigma}^{\phantom{\dagger}}$, since one can immediately see that $[\hat{N}_\sigma,\hat{\mathcal{H}}]=0$ and $[\hat{N}_\sigma,\hat{\mathcal{H}}_{ep}]=0$. These two operators can be combined into the $z$-component of two $SU(2)$ Lie algebras---a real spin algebra and a pseudospin algebra. The operators of the spin algebra are
\begin{equation}
    \hat{S}^z=\frac{1}{2}\sum_{x\in\Lambda}(\hat{n}_{x\uparrow}-\hat{n}_{x\downarrow}),~~\hat{S}^+=\sum_{x\in\Lambda}\hat{c}_{x\uparrow}^\dagger\hat{c}_{x\downarrow}^{\phantom{\dagger}},~~\text{and}~~\hat{S}^-=\sum_{x\in\Lambda}\hat{c}_{x\downarrow}^\dagger\hat{c}_{x\uparrow}^{\phantom{\dagger}}.
\end{equation}
One can immediately verify that they satisfy the standard $SU(2)$ algebra given by $\left [\hat{S}^+,\hat{S}^-\right ]=2\hat{S}^z$ and $\left [\hat{S}^{\pm},\hat{S}^z\right ]=\pm\hat{S}^{\pm}$ and that $\left [\hat{S}^{\pm},\hat{\mathcal{H}}\right ]=0$ and $\left [\hat{S}^{\pm},\hat{\mathcal{H}}_{ep}\right ]=0$, so that energy eigenstates of each Hamiltonian can be labeled by the total spin $s$ and the $z$-component of spin $m$ quantum numbers as well. The operators of the pseudospin algebra are
\begin{equation}
    \hat{J}^z=\frac{1}{2}\sum_{x\in\Lambda}(\hat{n}_{x\uparrow}+\hat{n}_{x\downarrow})-\frac{|\Lambda|}{2},~~\hat{J}^+=\sum_{x\in\Lambda}(-1)^{\epsilon(x)}\hat{c}^\dagger_{x\uparrow}\hat{c}^{\phantom{\dagger}}_{x\downarrow},~~\text{and}~~\hat{J}^-=\sum_{x\in\Lambda}\hat{c}_{x\downarrow}^{\phantom{\dagger}}\hat{c}_{x\uparrow}^{\phantom{\dagger}}.
\end{equation}
These operators also satisfy $\left [\hat{J}^+,\hat{J}^-\right ]=2\hat{J}^z$ and $\left [\hat{J}^z,\hat{J}^{\pm}\right ]=\pm\hat{J}^{\pm}$. In this definition, we must have that the lattice is bipartite, meaning the set of points $\Lambda=\Lambda_A\cup\Lambda_B$ (with $\Lambda_A\cap\Lambda_B=\varnothing$) and $t_{xy}=0$ if $x,y\in\Lambda_A$ or $x,y\in\Lambda_B$. The symbol $\epsilon(x)=0$ if $x\in\Lambda_A$ and $\epsilon(x)=1$ if $x\in\Lambda_B$. So the pseudospin operators can only be defined when the lattice for the hopping is bipartite. Then, it is only a symmetry of the Hubbard Hamiltonian if we also have that the coupling $U_x$ is independent of $x$, so we only discuss pseudospin symmetry for the Hubbard model on a bipartite lattice with spatially homogeneous interaction. In this case, one finds that $\left [\hat{J}^z,\hat{\mathcal{H}}\right ]=0$ and $\left [\hat{J}^{\pm},\hat{\mathcal{H}}\right ]=\pm U \hat{J}^{\pm}$. The pseudospin raising and lowering operators also raise and lower the energy of the energy eigenstates by $U$. Nevertheless, the total pseudospin $j$ and the $z$-component of pseudospin $m_j$ are both good quantum numbers, and energy eigenstates can be labeled by them as well. This scenario is similar to that of the Zeeman effect on independent spins when placed in a magnetic field.

The proofs in these theorems make statements about the quantum numbers $s$ and $j$ (where relevant) for the ground-state, but these results can be extended, to make statements of spin and pseudospin towers, first discussed in the Lieb and Mattis work from the early 1960s~\cite{lieb-mattis} that showed a spin tower for one-dimensional systems with nonsingular potentials (and hence, a lack of ferromagnetism). In a spin tower, the minimal energy state with spin quantum number $s$ lies strictly below the minimal energy state with spin quantum number $s+1$. We will discuss this more later in the paper. As a final note, our goal in this work is more to provide sketches of proofs and heuristic discussions---complete  rigorous proofs appear in the original literature.

\section{The two-site Hubbard model}

We now discuss a simple example, to be concrete about how quantum numbers enter into the models.
As an example for how the quantum numbers behave for the Hubbard model, we examine the two-site case, which can be diagonalized analytically.

The two-site ($|\Lambda|=2$) Hubbard model we will examine is 
\begin{equation}
    \hat{\mathcal{H}}=-t\sum_{\sigma}\left (\hat{c}_{1\sigma}^{\dagger}\hat{c}_{2\sigma}^{\phantom{\dagger}}+\hat{c}_{2\sigma}^{\dagger}\hat{c}_{1\sigma}^{\phantom{\dagger}}\right)+U\left (\hat{n}_{1\uparrow}^{\phantom{\dagger}}\hat{n}_{1\downarrow}^{\phantom{\dagger}}+\hat{n}_{2\uparrow}^{\phantom{\dagger}}\hat{n}_{2\downarrow}^{\phantom{\dagger}}\right ),
\end{equation}
with a uniform interaction $U$ and a hopping between the two sites. This is a bipartite lattice and it satisfies the uniformity condition on the interaction, so it has both spin and pseudospin symmetry.

In general, a lattice with $|\Lambda|$ sites has $4^{|\Lambda|}$ possible electronic states, because each site can have the state $0$, $\uparrow$, $\downarrow$, or $\uparrow\downarrow$ on it. Then, if we have $N_e$ total electrons in the quantum state, there are $\begin{pmatrix}2|\Lambda|\\N_e\end{pmatrix}=\frac{(2|\Lambda|)!}{N_e!(2|\Lambda|-N_e)!}$ possible states (which can be seen by filling the $2|\Lambda|$ spin orbitals with the $N_e$ electrons). So, for the two-site case, we have
\begin{align}
    N_e&=0,~~~ \begin{pmatrix}4\\0\end{pmatrix}=1~\text{state}\nonumber\\
    N_e&=1,~~~ \begin{pmatrix}4\\1\end{pmatrix}=4~\text{states}\nonumber\\
    N_e&=2,~~~ \begin{pmatrix}4\\2\end{pmatrix}=6~\text{states}\nonumber\\
    N_e&=3,~~~ \begin{pmatrix}4\\3\end{pmatrix}=4~\text{states}\nonumber\\
    N_e&=4,~~~ \begin{pmatrix}4\\4\end{pmatrix}=1~\text{state}\nonumber\\
    &~~~~~~~~~~~~~~~~~\,=16~\text{states total}~=4^2~\checkmark
\end{align}
When we have $N_e$ electrons, the maximal spin quantum number is $\frac{N_e}{2}$ and the minimum is $0$ or $\frac{1}{2}$ depending on whether $N_e$ is even, or odd, respectively. The allowed pseudospin quantum number  $m_j=\frac{N_e}{2}-\frac{|\Lambda|}{2}$, which means the pseudospin quantum number $j$ ranges in integer steps from $\left |\frac{N_e}{2}-\frac{|\Lambda|}{2}\right |$ up to either $\frac{|\Lambda|}{2}$ for $N_e$ even or $\frac{|\Lambda|-1}{2}$ for $N_e$ odd.

Before discussing the different energy eigenstates and their spin (and pseudospin) quantum numbers, we want to discuss some notation. The state with no electrons is denoted $|0\rangle$ and is called the vacuum state. Product states with $N_e$ electrons can be written as $\hat{c}_{x_1\sigma_1}^\dagger\cdots\hat{c}_{x_{N_e}\sigma_{N_e}}^\dagger|0\rangle$, which we rewrite in a simpler notation as $x_1\sigma_1x_2\sigma_2\cdots$ $x_{N_e}\sigma_{N_e}$ for efficiency. Our convention is that we order the site labels $x$ and we have $x_m\le x_n$ if $m<n$ and if two indices $x_m$ and $x_{m+1}$ are equal, the $\sigma_m=\uparrow$ appears before the $\sigma_{m+1}\downarrow$ index. In this work, we focus on using a real-space representation of the product states that we take linear combinations of to construct the many-body energy eigenstates (in particular, the ground state). Now we are ready to discuss the solutions of the Hubbard model with $|\Lambda|=2$. 

The state with $N_e=0$ is the vacuum state and it has $s=0$, $m=0$, $j=1$, and $m_j=-1$. Its energy is $E=0$. For $N_e=1$, we have four possible states, given by $1\uparrow$, $2\uparrow$, $1\downarrow$, and $2\downarrow$. The quantum numbers are $s=\frac{1}{2}$, $m=\pm\frac{1}{2}$, $j=\frac{1}{2}$ and $m_j=-\frac{1}{2}$. The energy eigenstates are 
\begin{equation}
    \frac{1}{\sqrt{2}}(1\sigma+2\sigma), ~E=-t~~\text{and}~~\frac{1}{\sqrt{2}}(1\sigma-2\sigma),~E=t.
\end{equation}
In both cases we can have $\sigma=\uparrow$ or $\downarrow$, so each state is two-fold degenerate. One can verify the energies (and that these are the eigenstates) by simply acting the Hamiltonian onto these states---for $N_e=1$, we only have the hopping term contributing.  For $N_e=2$, we have six possible states. We have the state with $s=0$, $m=0$, $j=1$, and $m_j=0$, found by applying $\hat{J}^+$ onto $|0\rangle$ and given by $\frac{1}{\sqrt{2}}(1\uparrow 1\downarrow-2\uparrow 2\downarrow)$. Its energy is raised by $U$ from the $N_e=0$ state energy, so it is equal to $U$. Next, we have the three degenerate states with $s=1$, $m=1,0,-1$, $j=0$, and $m_j=0$. They are given by $1\uparrow 2\uparrow$ ($m=1$), $\frac{1}{\sqrt{2}}(1\uparrow2\downarrow+1\downarrow 2\uparrow)$ ($m=0$), and $1\downarrow 2\downarrow$ ($m=-1$). The energy is $E=0$ (which is easiest to see by acting the Hamiltonian onto the $m=1$ state). Finally, we have two states with $s=0$, $m=0$, $j=0$, and $m_j=0$, spanned by $\frac{1}{\sqrt{2}}(1\uparrow2\downarrow-1\downarrow 2\uparrow)$ and $\frac{1}{\sqrt{2}}(1\uparrow1\downarrow-2\uparrow 2\downarrow)$. The Hamiltonian matrix that needs to be diagonalized to determine the energies is a $2\times 2$ matrix. It is constructed by simply working in the above basis, and is given by 
\begin{equation}
    \mathcal{H}_{s=0,m=0,j=0,m_j=0}=\begin{pmatrix}U&-2t\\-2t&0\end{pmatrix}.
\end{equation}
The energy eigenvalues are $E_\pm=\frac{U}{2}\pm\frac{1}{2}\sqrt{U^2+16t^2}$---the ground state corresponds to the minus sign for all $U$. The cases with $N_e=3$ and $N_e=4$ can be worked out by applying the $\hat{J}^+$ operator onto the $j=1$ and $j=\frac{1}{2}$ states. We have four states with $s=\frac{1}{2}$, $m=\pm\frac{1}{2}$, $j=\frac{1}{2}$ and $m_j=\frac{1}{2}$. They split into two two-fold degenerate multiplets, with energy given by $E=U\pm 2t$. Finally, the case with $N_e=4$ has $s=0$, $m=0$, $j=1$, and $m_j=1$. It has energy $2U$ and the state is $1\uparrow1\downarrow 2\uparrow 2\downarrow$.

We summarize these results as follows:
\begin{align}
    N_e&=0:~s=0,~m=0,~j=1,~m_j=-1,~~~~E=0~&1~\text{state}~\,\nonumber\\
    N_e&=1:~s=\frac{1}{2},~m=\pm\frac{1}{2},~j=\frac{1}{2}~,m_j=-\frac{1}{2},~~~~E=\pm t~\text{(twofold)}~&4~\text{states}\nonumber\\
    N_e&=2:~s=0,m=0,j=1,m_j=0,~~~~E=U~&1~\text{state}~\,\nonumber\\
    &~~~~~~~~~~~s=1,~m=-1,0,1~j=0~,m_j=0,~~~~E=0~\text{(threefold)}~&3~\text{states}\nonumber\\
    &~~~~~~~~~~~s=0,~m=0,~j=0,~m_j=0,~~~~E=\frac{U}{2}\pm\frac{1}{2}\sqrt{U^2+16t^2}~&2~\text{states}\nonumber\\
    N_e&=3:~s=\frac{1}{2},~m=\pm\frac{1}{2},~j=\frac{1}{2}~,m_j=\frac{1}{2},~~~~E=U\pm t~\text{(twofold)}~&4~\text{states}\nonumber\\
    N_e&=4:~s=0,~m=0,~j=1,~m_j=1,~~~~E=2U~&1~\text{state}~.
\end{align}
What one can see from this is that the ground state always has minimal spin and minimal pseudospin; it even has both spin and pseudospin towers at half filling. In general, Lieb proved that the ground state for the attractive case has minimal spin and for the repulsive case has minimal pseudospin; in both cases, the ground state is also unique, with additional restrictions on the problem. But his proofs show even more than this as well when combined with the partial particle-hole transformation and when one explores other consequences of these ideas.

\section{Spin-reflection positivity and the proof of the existence of a spin-singlet ground state}

We will follow Ref.~\cite{freericks-lieb} for both proofs, because it is simpler to discuss both at the same time using the language developed in the later paper. We will discuss preliminaries for the models and then state and prove the first theorem (existence of a spin singlet ground state) for both models. In the next section, we will discuss and sketch the proof for the uniqueness of the ground state.

In the existence proof, we will work with a concrete representation of the many-body wavefunction. The first requirement we have is that the number of electrons is even and that the number of lattice sites is finite. Next, we restrict the many-body states to $m=0$, because states with every possible $s$ eigenvalue have a representative in this subspace. Then, we have the same number of up-spin and down-spin electrons, which we denote $N$ and hence $N_e=N_\uparrow+N_\downarrow=N+N=2N$. It is important to note as well that the two Hamiltonians we consider are real and are unchanged if we interchange the up spins with the down spins and vice versa. This is where the requirement that the hopping matrix be real, that the hopping not depend on spin, and that the electron-phonon coupling be too the total electronic charge at a given lattice site are required. This means, if I have a wavefunction that corresponds to an energy eigenstate, I can interchange the spins, or I can take its complex conjugate, or do both, and it will still be an energy eigenstate. We will use this important observation in just a moment. We will use a first-quantized notation for our description.

\textit{Theorem 1 (Existence of a spin-singlet ground state for the attractive Hubbard model in Eq.~(\ref{eq:hubbard}))}: If the interactions are attractive, so that $U_x\le 0~\forall~x\in\Lambda$, then amongst all of the possible ground-states of the Hubbard model with an even number of electrons $N_e=2N$, there is at least one state that is a spin-singlet state with $s=0$.

We now discuss the proof of Theorem 1 for the Hubbard model, where we have $U_x\le 0,~\forall x\in\Lambda$, so that we can write them as $U_x=-|U_x|$. We let $X=(x_1,x_2,\cdots,x_N)$ be the $N$-tuple of labels for the up-spin electrons and $Y=(y_1,y_2,\cdots,y_N)$ be the corresponding down-spin labels. We let $\Psi(X,Y)$ denote the many-body wavefunction---it must be separately antisymmetric under an interchange of any of the $x_i$ coordinates and under an interchange of any of the $y_i$ coordinates. Because $\Psi(Y,X)$,  $\Psi^*(X,Y)$, and $\Psi(Y,X)^*$ are also eigenstates with the same energy, we can, without loss of generality, assume that $\Psi(X,Y)=\Psi(Y,X)^*$---the wavefunction is self-adjoint, when thought of as a matrix. This is because $\Psi(X,Y)+\Psi(Y,X)^*$ and $i(\Psi(X,Y)-\Psi(Y,X)^*)$ are both solutions as well. Note that this assumption does not imply that $\Psi(X,Y)$ is real valued. This matrix $\Psi$ is a $d\times d$ matrix, with $d=\begin{pmatrix}|\Lambda|\\N\end{pmatrix}$. We use the spectral theorem, which says that any finite-dimensional self-adjoint matrix can be expanded in an eigenfunction basis $\{\phi_\alpha\mid \sum_Y\Psi(X,Y)\phi_\alpha(Y)=w_\alpha\phi_\alpha(X)\}$, with real eigenvalues $w_\alpha$. Note that each $\phi_\alpha$ is orthonormal, with $\sum_X \phi_\alpha^*(X)\phi_\beta(X)=\delta_{\alpha\beta}$ and that they are antisymmetric functions with respect to interchange of any two of the $x_i$ labels. The expansion for $\Psi(X,Y)$ becomes
\begin{equation}
    \Psi(X,Y)=\sum_{\alpha=1}^d w_\alpha\phi_\alpha(X)\phi_\alpha^*(Y).
    \label{eq:spectral}
\end{equation}

We now compute the expectation value of the energy, in the state $\Psi$ using the spectral decomposition in Eq.~(\ref{eq:spectral}). The kinetic-energy piece is 
\begin{align}
    \langle\Psi|\hat{K}|\Psi\rangle&=\sum_{\alpha=1}^d\sum_{\beta=1}^dw_\alpha w_\beta\sum_X\sum_Y\sum_Z\Big(\phi_\alpha^*(X)K_\uparrow(X,Y)\phi_\beta(Y)\phi_\alpha(Z)\phi^*_\beta(Z)\nonumber\\
    &+\phi_\alpha(X)K_\downarrow(X,Y)\phi_\beta^*(Y)\phi_\alpha^*(Z)\phi_\beta(Z)\Big).
\end{align}
Here, $\hat{K}_\sigma(X,Y)$ is the first-quantized version of $\sum_{x,y\in\Lambda}t_{xy}\hat{c}_{x\sigma}^\dagger\hat{c}_{y\sigma}^{\phantom{\dagger}}$, expressed in the same coordinate system as we used for the wavefunctions. Using the facts that the $\{\phi_\alpha\}$ are an orthonormal set, that the first quantized form $K_\sigma$ is independent of spin, and that $t_{xy}$ are real, yields
\begin{equation}
    \langle\Psi|\hat{K}|\Psi\rangle=2\sum_{\alpha=1}^dw_\alpha^2\langle\phi_\alpha|\hat{K}|\phi_\alpha\rangle.
\end{equation}
Similarly, we have the potential energy piece satisfies
\begin{align}
    \langle \Psi|\hat{L}|\Psi\rangle=-\sum_{x\in\Lambda}|U_x|\sum_{\alpha=1}^d\sum_{\beta=1}^dw_\alpha w_\beta\sum_W\sum_X\sum_Y\sum_Z&\phi_\alpha^*(W)L_x(W,X)\phi_\beta(X)\nonumber\\
    &\times\phi_\alpha(Y)L_x(Y,Z)\phi_\beta^*(Z),
\end{align}
where $L_x(X,Y)$ is the first quantized version of $n_{x\sigma}$, which is independent of $\sigma$. Using the fact that $L_x^*(X,Y)=L_x(X,Y)$ then yields
\begin{equation}
    \langle \Psi|\hat{L}|\Psi\rangle=-\sum_{x\in\Lambda}|U_x|\sum_{\alpha=1}^d\sum_{\beta=1}^dw_\alpha w_\beta|\langle\phi_\alpha|\hat{L}_x|\phi_\beta\rangle|^2.
\end{equation}
Finally, the overlap of the wavefunction with itself is
\begin{equation}
    \langle\Psi|\Psi\rangle=\sum_{\alpha=1}^d\sum_{\beta=1}^d\sum_X\sum_Y\phi_\alpha^*(X)\phi_\beta(X)\phi_\alpha(Y)\phi_\beta^*(Y)=\sum_{\alpha=1}^dw_\alpha^2.
\end{equation}
This leads to the energy expectation value
\begin{equation}
    E(\Psi)=\frac{2\sum_{\alpha=1}^dw_\alpha^2\langle\phi_\alpha|\hat{K}|\phi_\alpha\rangle-\sum_{x\in\Lambda}|U_x|\sum_{\alpha=1}^d\sum_{\beta=1}^dw_\alpha w_\beta|\langle\phi_\alpha|\hat{L}_x|\phi_\beta\rangle|^2}{\sum_{\alpha=1}^dw_\alpha^2}.
\end{equation}
One can immediately see that replacing $w_\alpha$ by $|w_\alpha|$ can only make the energy smaller, so we can assume that the ground state includes a state that satisfies $\Psi=|\Psi|$ in a matrix sense, where $|\Psi|=\sqrt{\Psi^2}$; note that this is usually not the same as $\Psi(X,Y)\to|\Psi(X,Y)|$, as that only holds in the diagonal basis. This means that $\text{Tr}\,\Psi=\sum_{\alpha=1}^d|w_\alpha|>0$, which implies that $\Psi(X_0,X_0)\ne 0$ for some $X_0$. Then the up spin particles are at the same locations as the down spin particles---this requires them to be in a spin-singlet state. So, amongst the possible ground states, there is at least one ground state that is a spin singlet.

\textit{Theorem 2 (Existence of a spin-singlet ground state for the generalized electron-phonon Hamiltonian in Eq.~(\ref{eq:el-ph}))}: If the boundedness criterion in Eq.~(\ref{eq:bounded}) holds, then among the different ground states of the generalized electron-phonon Hamiltonian, there is at least one ground state that has $s=0$.

The proof of Theorem 2 for the electron-phonon Hamiltonian is similar. Again, we work in the space with $m=0$ and we have a $\vec{q}$-valued $d\times d$ matrix, with the same $d$ as before, when we have a lattice with $|\Lambda|$ sites and $N$ spin up and $N$ spin down electrons. As before, taking complex conjugates and transposes, we can show that for each point $\vec{q}$, the matrix $\Psi(\vec{q})$ is a self-adjoint matrix for each $\vec{q}$. Recall as well that we have $\nu$ different phonon modes (we do not require $|\Lambda|=\nu$). We write the Schr\"odinger equation in a schematic form, where we separate out the kinetic-energy contribution of the phonons from the rest of the potential (which we call $\hat{V}(\vec{q})$---it is a self-adjoint matrix for each $\vec{q}$). The potential acts on the wavefunction matrix from the right and from the left, to incorporate both the up-spin contribution and the down-spin contribution. The schematic Schr\"odinger equation $\hat{\mathcal{H}}_{ep}\Psi=E\Psi$ takes the following form:
\begin{equation}
    -\sum_{i=1}^\nu\frac{\hbar^2}{2m_i}\frac{\partial^2}{\partial q_i^2}\Psi(\vec{q})+V(\vec{q})\Psi(\vec{q})+\Psi(\vec{q})V(\vec{q})=E\Psi(\vec{q}),
\end{equation}
where $E$ is the energy eigenvalue.

We will use the Schr\"odinger equation to determine the expectation value of the energy. But first, let's review how this calculation works. Since $\Psi(\vec{q})$ is a self-adjoint matrix, we can again write it in a diagonal form for each $\vec{q}$, using the ``instantaneous'' eigenvectors as
\begin{equation}
    \Psi(X,Y;\vec{q})=\sum_{\alpha=1}^dw_\alpha(\vec{q})\phi_\alpha(X;\vec{q})\phi_\alpha^*(Y;\vec{q}).
\end{equation}
We use the symbol $\mathbb{R}^\nu$ to denote the position space for the phonons, which is $\nu$-dimensional and $d\vec{q}$ denotes the integration measure for the $\nu$-dimensional space.
The orthogonal eigenvectors satisfy $\sum_X\phi_\alpha^*(X;\vec{q})\phi_\beta(X;\vec{q})=\delta_{\alpha\beta}$ and are antisymmetric functions with respect to the interchange of any two of the $x_i$ labels. Then, the norm of the state $|\Psi\rangle$ is
\begin{align}
    \langle\Psi|\Psi\rangle&=\int_{\mathbb{R}^\nu}d\vec{q}\sum_{\alpha=1}^d\sum_X\sum_Yw_\alpha(\vec{q})w_\beta(\vec{q})\phi_\alpha^*(X;\vec{q})\phi_\beta(X;\vec{q})\phi_\alpha(Y;\vec{q})\phi_\beta^*(Y;\vec{q})\nonumber\\
    &=\int_{\mathbb{R}^\nu}d\vec{q}\sum_{\alpha=1}^dw_\alpha^2(\vec{q}).
\end{align}
Using a similar methodology, we find that the numerator of the energy expectation value becomes
\begin{equation}
    \langle \Psi|\mathcal{H}_{ep}|\Psi\rangle=\int_{\mathbb{R}^\nu}d\vec{q}\left (\sum_{i=1}^\nu\frac{\hbar^2}{2m_i}\text{Tr}\,\left [\frac{\partial}{\partial q_i}\Psi(\vec{q})\right ]^2+2\text{Tr}\,\left [ V(\vec{q})\Psi^2(\vec{q})\right ]\right ).
\end{equation}
Note that the trace and the matrix multiplications here are with respect to the $d\times d$ matrix structure of the respective terms.

Our strategy is the same as before. We replace $\Psi(\vec{q})$ by $|\Psi(\vec{q})|=\sqrt{\Psi(\vec{q})^2}$. Clearly the norm of the state and the potential-energy term are unchanged by this replacement. Proving that the phonon kinetic energy is also not increased is a more complicated technical question involving distributions, but all of those details are fully handled in the original publication~\cite{freericks-lieb}, so indeed, we have that the energy is not increased. Hence, among the ground states, there always is one that satisfies $\Psi(\vec{q})=|\Psi(\vec{q})|$. Using the same reasoning as for the Hubbard-model proof, one then concludes that the ground state must include an $s=0$ state.

\section{Sketch of the uniqueness proofs}

The existence proofs in the previous subsection illustrated the concept of spin reflection positivity. When we wrote the wavefunction in the square matrix form (possible because we are in the sector with $m=0$), we made manifest the spin-reflection symmetry, which switches the up spins and the down spins and vice-versa, and showed that a positivity criterion for the wavefunction, does not raise the variational energy, hence the ground state must be in this positive semidefinite form. From this we learn that the ground state has a $s=0$ component to it. 

The existence proofs are both technically more complicated and they require significant additional restrictions on the Hamiltonians. We will not prove them completely here, but we will sketch how the arguments go for these theorems. Key to this work is the idea of a lattice structure in the many-body configuration space of the wavefunction. If this configuration space is fully connected, then the ground-state is unique (similar in many respects to the methodology used to prove the uniqueness in the Perron-Frobenius theorem, which also relies on positivity and connectivity arguments). 

The additional restrictions for the Hubbard model are not too severe---we must have the hopping be connected on the graph $\Lambda$ and we must require all $U_x$ to be strictly negative $U_x<0$. The concept of connected on the lattice $\Lambda$ is defined as follows: every two lattice sites $x,y\in\Lambda$ are said to have a bond between them if $t_{xy}\ne 0$. Then, the lattice is connected, if for every pair $x\ne y\in\Lambda$, there is a chain of bonds along a path that connects $x$ to $y$. This path (with the chain of bonds) is defined as $\{x_i\mid x_i\in\Lambda,x_i\ne x_j~\text{if}~i\ne j,~ x=x_1,\,x_2,\,\cdots,\,x_{n-1},\,x_n=y,~\text{and}~t_{x_1x_2}t_{x_2x_3}\cdots t_{x_{n-1}x_n}\ne 0\}$. Hence, we say for every pair of lattice points $x$ and $y$ there is a path from $x$ to $y$ connected by a chain of bonds. This is the conventional condition for connectedness of a lattice via its bonds.

The lemma we will need to prove first is that if the lattice $\Lambda$ is connected, then the single-spin many-body configuration space is also connected, when thought of as an abstract lattice in the many-body space. The single-spin many-body configuration space is the collection of all many-body states with $N$ electrons of the form
\begin{equation}
    \{\hat{c}_{x_1}^\dagger\cdots\hat{c}^\dagger_{x_N}|0\rangle\mid x_i\in\Lambda~\forall~i=1,\cdots,N;~x_i\ne x_j ~\forall~ i,j=1,\cdots,N~\text{when}~ i\ne j\},
    \label{eq:many-body}
\end{equation}
in the second-quantization language. Note that we have suppressed the spin label here. You can think of these as exclusively spin-up or exclusively spin down electron states. We can collect all of many-body states into a many-body graph, denoted $\tilde\Lambda$, defined by the collection of all of the many-body states, as defined in Eq.~(\ref{eq:many-body}). We index these states with the same $N$-tuple as we used before, $X=(x_1,x_2,\cdots,x_N)$, with all $x_i\in\Lambda$. We say that two many-body states are connected by a many-body bond on $\tilde\Lambda$ if the two graph states $X$ and $Y$ satisfy
\begin{equation}
    X=(x_1,x_2,\cdots,x,\cdots,x_N)~~\text{and}~~Y=(x_1,x_2,\cdots,y,\cdots,x_N)
\end{equation}
where the two $N$-tuples are identical except for one element---$X$ contains the element $x$, while $Y$ contains the element $y$ with all other indices identical---and $t_{xy}\ne 0$. Heuristically, this means the many-body state $Y$ can be reached from the many-body state $X$ by acting the kinetic-energy operator $\hat{K}$ on it.

We also need to define the one-dimensional projector $\hat{\Pi}^X=\hat{n}_{x_1}\hat{n}_{x_2}\cdots\hat{n}_{x_N}$ (in the second-quantized language) as the projector onto the the many-body state, where electrons are placed at the $N$-lattice sites (in $\Lambda$) and labeled by the $N$-tuple $X$. It is easy to see that this is a Hermitian operator and that it is a one-dimensional projector (because $\hat{n}_x^2=\hat{n}_x$). Note that because number operators commute with each other, the projector is identical for two states $X$ and $X'$ that are just a permutation of the respective indices. It is therefore useful to work, instead with the equivalence classes of the projectors, with respect to the permutation operation.

\textit{Lemma 1 (Many-body connectedness of the single-spin configuration space lattice $\tilde\Lambda$)}: If the $\Lambda$ lattice is connected, such that any two points $x,y\in\Lambda$ can be connected by a path in $\Lambda$ such that a bond connects each step along the path with a nonzero hopping matrix element, then the many-body lattice $\tilde\Lambda$ is also connected with respect to the kinetic energy operator, in the sense that for any $X,Y\in\tilde\Lambda$, there exists a set of elements $X_i\in\tilde\Lambda$ with $X=X_1,X_2,\cdots,X_{m-1},X_m=Y$ such that $X_i\ne X_j$ for all $i\ne j$ and the product of matrix elements, determined by the operator chain below satisfies
\begin{equation}
    \hat{\Pi}^{X_m}\hat{K}\hat{\Pi}^{X_{m-1}}\cdots\hat{\Pi}^{x_{2}}\hat{K}\hat{\Pi}^{X_1}\ne 0.
    \label{eq:connectivity-op}
\end{equation}

\begin{figure}
    \centering
    \includegraphics[width=0.9\textwidth]{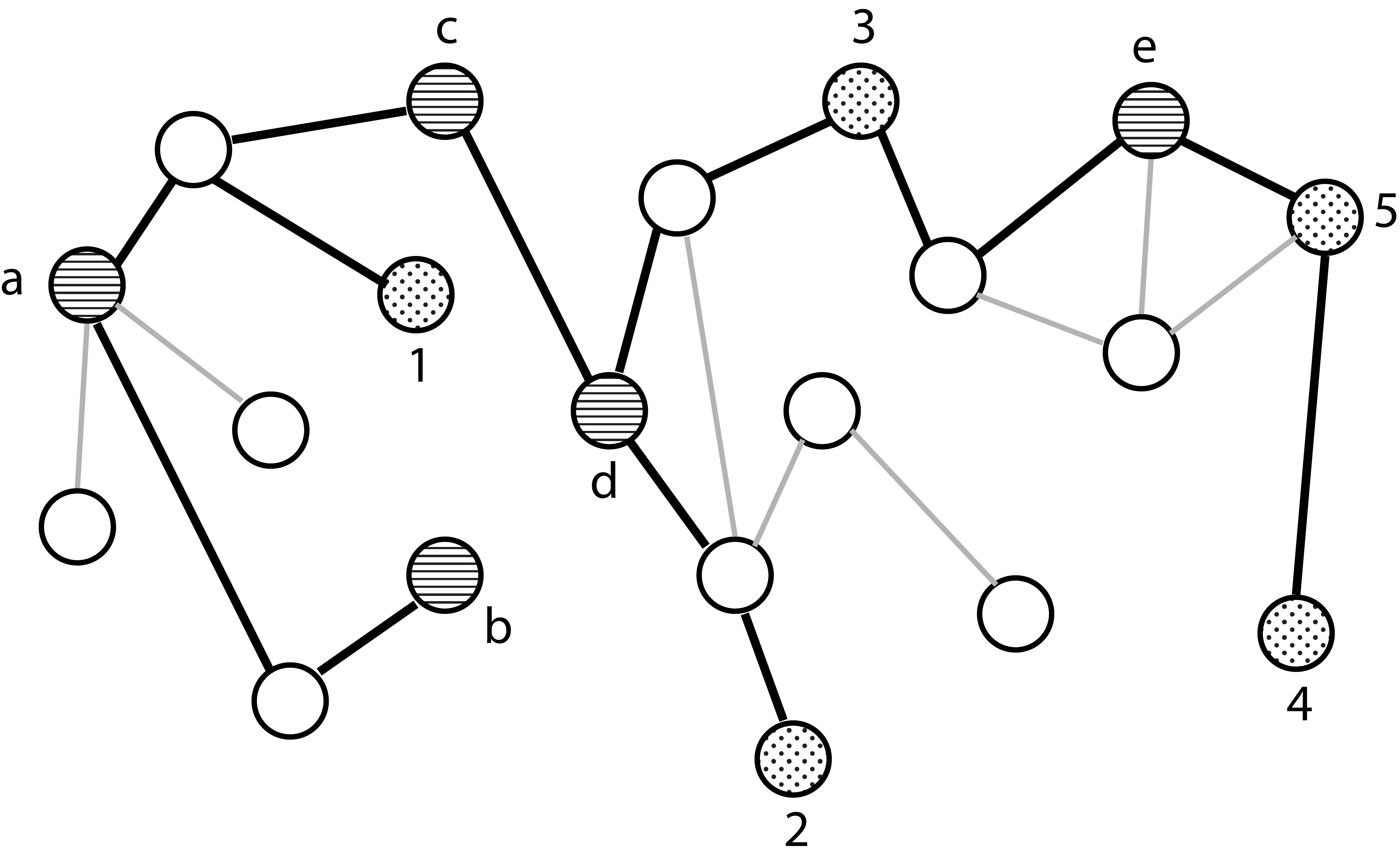}
    \caption{Schematic for the proof of the connectivity of $\tilde\Lambda$. The figure illustrates an initial state $X$ (circles with a dot pattern) and a final state $Y$ (circles with a horizontal line pattern), which are elements of $\tilde\Lambda$. Here, the circles that are filled with a pattern illustrate the initial state (with electrons on sites 1--5) and the final state (with electrons on sites a--e). The lines illustrate the bonds between the sites on the original lattice $\Lambda$. The solid black lines indicate the bonds in $\Lambda$ over which we will be moving electrons from one site to another. Each intermediate state in this figure corresponds to a particular state $\tilde X$ on the many-body configuration lattice $\tilde\Lambda$. The lattice $\tilde\Lambda$ is not illustrated here.}
    \label{fig:lambda}
\end{figure}

The proof begins with solving a geometrical problem, which is visualized concretely in Fig.~\ref{fig:lambda}. We put unlabeled markers on the lattice $\Lambda$ at the $N$ locations of the initial configuration $X=(x_1,\cdots,x_N)\in\tilde\Lambda$. Our goal is to move those markers, via bonds in $\Lambda$ (determined by nonzero hopping matrix elements connecting two lattice sites) from the sites corresponding to configuration $X$ to the sites corresponding to the configuration $Y$. This is illustrated in the figure as follows: (i) sites corresponding to $X$ are the circles with a dotted pattern, with the labels indicated by numerals from 1 to 5; (ii) sites corresponding to $Y$ are the circles with a horizontal line pattern, with the labels indicated by letters from a to e; and (iii) solid dark lines indicate the bonds we will use to move the markers during our algorithm. We first describe how it works for the concrete figure, and then describe the general case.

In our first step, we will move a marker from site 1 to site a, along the black-line bonds, to end in a state with one marker on a and four markers on 2--5. This first step is simple, as we have a direct path from 1 to a with no other markers along the path, so we make the move. Next, we want to move the marker from 2 to b. But here, the marker on a blocks our ability to move the marker from 2 to b. So, we move the marker from 2 to the empty circle adjacent to a, then move the marker from a to b (which has a direct path) and then finally move the marker from the site adjacent to a to a. Now we have markers on a and b and 3--5. Moving the marker from 3 to c is direct, with no obstructions, so we do it. Moving from 4 to d is blocked on the first attempt. Instead, we move from 5 to d, and then from 4 to 5. Finally, in the last step, we move from 5 to e. As you can see, when we move the markers from 1--5 to a--e, we can always succeed, but we do not always move the marker originally sitting on 1 to site a and similarly for the remaining markers---when there is blockage, we must move them in tandem---but this is always possible due to the connectivity of the original lattice $\Lambda$. This shows $X$ and $Y$ are connected on $\tilde\Lambda$.

How about the general case? We proceed in essentially the same way. Start by placing markers on the $N$ sites given by $X=(x_1,x_2,\cdots,x_N)$. Next, find a path from the site $x_1$ to the site $y_1$. Such a path is guaranteed by the connectivity of $\Lambda$. Identify whether any other markers are on the path. If none, move the marker directly from $x_1$ to $y_1$. If there are markers along the path, then move the marker that has a direct path to $y_1$ from its lattice site to $y_1$. Then move the next marker down the path (marching backwards from $y_1$ towards $x_1$) from where it sits to where the marker originally was that we moved to $y_1$. Repeat this process for all remaining markers on the path. In this fashion, we have moved a marker from $x_1$ to $y_1$, and have left unchanged the markers on sites $x_2$ to $x_N$. Moving the marker from $x_2$ to $y_2$ proceeds in a similar fashion. Note that markers along the path could be on an $X$ site or on a $Y$ site. It does not matter, the algorithm works for either scenario, as you can readily check.

Now, with the geometrical problem solved, we select the $m$ $X$ configurations $X_1,\cdots,X_m$ by choosing every configuration of the markers that were used in the rearrangement to move the markers from $X$ to $Y$. Because $\hat{\Pi}^{X_j}$ is a projection operator, so $\hat{\Pi}^{X_j}=\left (\hat{\Pi}^{X_j}\right )^2$, we can rewrite
\begin{equation}
    \hat{\Pi}^{X_m}\hat{K}\hat{\Pi}^{X_{m-1}}\cdots\hat{\Pi}^{x_{2}}\hat{K}\hat{\Pi}^{X_1}=\left (\hat{\Pi}^{X_m}\hat{K}\hat{\Pi}^{X_{m-1}}\right )\left (\hat{\Pi}^{X_{m-1}}\hat{K}\hat{\Pi}^{X_{m-1}}\right)\cdots\left (\hat{\Pi}^{x_{2}}\hat{K}\hat{\Pi}^{X_1}\right ).
\end{equation}
We now introduce complete sets of states in between each projection operator and to the left and to the right. The states $X_j$ and $X_{j+1}$ in $\tilde\Lambda$ agree on all but two of their index values (recall that we work with the equivalence class due to permutations of each $X$ state). This means we have $X_j=(x_1,x_2,\cdots,x,\cdots,x_N)$ and $X_{j+1}=(x_1,x_2,\cdots,y,\cdots,x_N)$. Then the matrix element satisfies
\begin{equation}
    \langle \chi_\alpha|\hat{\Pi}^{X_{j+1}}\hat{K}\hat{\Pi}^{X_j}|\phi_\beta\rangle=t_{xy}\phi_\alpha^*(X_{j+1})\chi_\beta(X_j)N!,
\end{equation}
where the states $\{\chi_\alpha\}$ are the localized states, formed by products of creation operators at each lattice site in $\Lambda$ acting on the vacuum. In this basis, only one state $\chi_\beta$ will yield a nonzero matrix element for a given $\chi_\alpha$ and the operator $\hat{\Pi}^{X_{j+1}}\hat{K}\hat{\Pi}^{X_j}$. Then, because of the connectivity of the original lattice $\Lambda$, we have that the product of matrix elements is nonzero, as we claimed. Hence, the state $X$ and $Y$ in $\tilde\Lambda$ are connected.

\textit{Theorem 3: (Uniqueness of the Hubbard model when $\Lambda$ is connected and $U_x<0~\forall~x\in\Lambda$)}: The ground state of the Hubbard model when the interactions are strictly attractive and the lattice is connected is unique. Since we always have a ground state with $s=0$, the ground state must have $s=0$ as well.

We sketch the uniqueness proof for the Hubbard model. The proof is by contradiction. Assume two solutions exist $\Psi_1$ and $\Psi_2$. Then the state $\Psi(\lambda)=\Psi_1+\lambda\Psi_2$ is also a ground state for all real $\lambda$. For some value of $\lambda$ the matrix $\Psi(\lambda)$ is neither positive semidefinite nor negative semidefinite. Fix $\lambda$ for this value, then since $|\Psi(\lambda)|$ is also a ground state, the two matrices $\Psi_\pm=\frac{1}{2}(|\psi(\lambda)|\pm\Psi(\lambda))$ are also ground states---the state $\Psi_+$ is a positive semidefinite ground state.

$\Psi_+$ satisfies the following Schr\"odinger equation:
\begin{equation}
    \hat{K}\Psi_++\Psi_+\hat{K}-\sum_{x\in\Lambda}|U_x|\hat{L}_x\Psi_+\hat{L}_x=E\Psi_+.
    \label{eq:schrodinger-hubbard}
\end{equation}
Define $H_+$ as the range of $\Psi_+$ and $H_\perp$ as the orthogonal complement of $H_+$ in the Hilbert space. If there are two ground states, then both $H_+$ and $H_\perp$ are nontrivial subspaces of the Hilbert space. If we multiply Eq.~(\ref{eq:schrodinger-hubbard}) on the right and the left by $\hat{\Pi}_\perp$, the projector onto $H_\perp$, we find
\begin{equation}
    \Psi_+\hat{L}_x\hat{\Pi}_\perp=0.
    \label{eq:proj1}
\end{equation}
This result follows because each $U_x\ne 0$, $\Psi_+\hat{\Pi}_\perp=\hat{\Pi}_\perp\Psi_+=0$, and $\Psi_+$ is a positive semidefinite matrix. This says that $\hat{L}_x$ maps $H_+$ to $H_+$. Keeping in mind Eq.~(\ref{eq:proj1}), we multiply the Schr\"odinger equation in Eq.~(\ref{eq:schrodinger-hubbard}) on the right by $\hat{\Pi}_\perp$ and find that
\begin{equation}
    \Psi_+\hat{K}\hat{\Pi}_\perp=0,
\end{equation}
as well. This means $\hat{K}$ also maps $H_+$ to $H_+$. Hence, the subspaces $H_+$ and $H_\perp$ are left invariant by $\hat{K}$ and $\hat{\Pi}_\perp$. 

The sum of the projectors, over one element of each equivalence class, is the identity element on the Hilbert space, so there is some $X_0\in\tilde\Lambda$ such that $\hat{\Pi}^{X_0}\Psi_+\ne0$ because $H_+$ is nontrivial. Furthermore, for some element $\psi_\perp\in H_\perp$, there exists a $Y_0\in\tilde\Lambda$ such that $\hat{\Pi}^{Y_0}\psi_\perp\ne 0$. Now, choose the operator in Eq.~(\ref{eq:connectivity-op}) with $X=X_0$ and $Y=Y_0$. We must have the matrix element between $\langle\psi_\perp|$ and $|\Psi_+\rangle$ is nonzero. But this operator keeps the subspaces $H_+$ and $H_\perp$ invariant, because it is constructed from $\hat{K}$ and $\hat{\Pi}^X$ operators. Hence, the matrix element cannot be nonzero. This is a contradiction. So, the ground state must be unique. Since it is unique, it must have $s=0$.

\textit{Theorem 4: (Uniqueness of the ground state for the electron-phonon Hamiltonian)}: When the electron-phonon Hamiltonian satisfies four more conditions: (i) the hopping matrix elements are independent of $\vec{q}$ and the lattice $\Lambda$ is connected; (ii) the remaining functions of $\vec{q}$ in the Hamiltonian are differentiable; (iii) the $G_x(\vec{q})$ functions are independent, meaning the $|\Lambda|\times\nu$ matrix $\partial G_x(\vec{q})/\partial q_j$ is rank $|\Lambda|$ for all $\vec{q}$; and (iv) all masses $m_j$ are finite, then the ground state is unique and has $s=0$.

We will not sketch the proof of this theorem here. It is closely related to the proof for the Hubbard model, but it requires many additional technical details to be properly handled. It is given in the original work~\cite{freericks-lieb}. So, we do not repeat it here.

While these proofs focused on electronic systems where either the direct interaction, or the effective electron-electron interaction, are all attractive, then one has the ground state is always $s=0$ and under some additional restrictions, it is unique. These theorems are reasonable physically---when the attraction is very strong, it is difficult to break a pair, so all electrons are paired, which must be a singlet state. In the limit as the interaction goes to zero, if the hopping matrix has no degeneracies, then we fill in the lowest energy levels, and this is also a singlet state (if we have degeneracies, which is usually the case, then the ground state includes a singlet state, but need not be unique). The main achievement of Lieb's work is to show the ground state is a spin singlet and is unique for all intermediate interactions.

\section{Impact of the original ideas}

It turns out that these original ideas can be generalized to other cases that are more physically relevant (the direct electron-electron interaction is usually repulsive, not attractive, for example). To understand this requires us to discuss the partial particle-hole transformation.

The partial particle-hole transformation is a particle-hole transformation on the up-spin particle, leaving the down spin particles unchanged; it requires we work on a bipartite lattice for the Hubbard model with $U$ independent of $x$. In this transformation, we have
\begin{equation}
    \hat{c}_{x\uparrow}^{\phantom{\dagger}}\to \hat{d}_{x\uparrow}^\dagger(-1)^{\epsilon(x)}~~\text{and}~~\hat{c}_{x\downarrow}^{\phantom{\dagger}}\to\hat{d}_{x\downarrow}^{\phantom{\dagger}}.
\end{equation}
In this case, the number operators transform as $\hat{n}_{x\uparrow}\to(1-n_{x\uparrow})$ and $\hat{n}_{x\downarrow}\to\hat{n}_{x\downarrow}$. Hence, the filling changes by $N_\uparrow\to|\Lambda|-N_\uparrow$ and $N_\downarrow\to N_\downarrow$. This means all even fillings are transformed to half filling $N_\uparrow+N_\downarrow=|\Lambda|$. Furthermore, the interaction changes sign (plus a shift in energy given by $UN_{\downarrow}$), so the attractive model becomes the repulsive model (the hopping term remains intact, if the lattice is bipartite).  Be-cause it is a unitary transformation, the energies remain the same, in particular, the repulsive Hubbard model, after the constant shift, has the same ordering of the energy levels as in the attractive model. But the spin operators transform to the pseudospin operators and the pseudospin operators transform to the spin operators. Key, is the fact that the ground state is unique, so it has no level crossings. This means its spin quantum number, which is heretofore unknown, cannot change with $U$. But, in the large-$U$ limit, the Hubbard model maps to a Heisenberg antiferromagnet. So, it must have the same spin quantum number as the antiferromagnet. But, this is known to be $\Big||\Lambda_A|-|\Lambda_B|\Big|/2$. So, in the case where the two sublattices of the bipartite lattice have different numbers of lattice sites, the system is ferrimagnetic (no long-range order was proven, just that the spin quantum number is in between the saturated ferromagnet and the spin singlet). Most bipartite lattices that people were familiar with at that time had $|\Lambda_A|=|\Lambda_B|$. Lieb pointed out that by removing bonds from the face-centered-cubic lattice, one could create a bipartite lattice with one sublattice having three-times as many sites as the other sublattice. These lattices are special and bipartite lattices with different numbers of sites on each lattice are now called Lieb lattices. They are characterized by having a macroscopic number of zero eigenvalues in the hopping matrix, and are often called flat band models. We will have more to say about them later. 

Because the Hubbard model is closely related to the periodic Anderson model (which can be thought of as a Hubbard model with two bands and $U=0$ in the second band, with the hybridization being thought of as a hopping term), the theorem Lieb proved holds for these models as well. This has led to work by Sigrist and collaborators on proving properties of the spin quantum numbers of this and related models~\cite{sigrist1,sigrist2}. Tian was even able to prove antiferromagnetic order~\cite{tian1}.

Another interesting area where significant work has been done is in the description of spin and pseudospin towers. A spin tower was proven to exist in one dimension in the original work by Lieb and Mattis in the 1960s~\cite{lieb-mattis}. The spin tower says the lowest energy eigenvalue with total spin quantum number $s+1$ is greater than the lowest energy eigenvalue with total spin quantum number $s$. The Lieb-Mattis result can only be proven in one dimension and with finite-strength potentials. Lieb's original proof for the Hubbard model~\cite{lieb-hubbard} shows that the ground-state for the attractive model has $s=0$ whenever the conditions for uniqueness hold. Noce and Romano~\cite{romano} showed that the one-dimensional attractive Hubbard model has minimal pseudospin in its ground state, which implies a pseudospin tower for all even fillings. Shen wrote two papers discussing these issues~\cite{shen1,shen2}. Shen was able to extend Lieb's proof, via a small change of the uniqueness proof, to also show that $|W|$ is positive definite, as opposed to positive semidefinite. This allows one to determine quantum numbers of the ground state if we can find a positive semidefinite state with a specific quantum number---such a state has nonzero overlap with the positive-definite state and so they must share the same eigenvalue. Shen employed this idea along with the ground states of a fictitious pseudospin model, to show that the ground states of the attractive Hubbard model have minimal pseudospin for $N\le 2\min\{|\Lambda_A|,|\Lambda_B|\}$, then the pseudospin is fixed at $\Big||\Lambda_A|-|\Lambda_B|\Big|/2$ until $N\ge 2\max\{|\Lambda_A|,|\Lambda_B|\}$, where it becomes minimal again. Boretsky and Freericks~\cite{freericks-tower} constructed the pseudospin states concretely. 

One can perform the partial particle-hole transformation, and then learn that the half-filled repulsive Hubbard model has a partial or a full spin tower---the tower forms for spin values that satisfy $s\ge \Big||\Lambda_A|-|\Lambda_B|\Big|/2$, because the ground state has $s=\Big||\Lambda_A|-|\Lambda_B|\Big|/2$. We do not learn about the relative ordering of lower spin values, except they are above the ground state in energy.

What remains unproven is whether there exists a spin tower for the attractive Hubbard model. Heuristically this should occur, as it is clear for $U\to-\infty$ and $U\to 0$, but is has not yet been proven. If it is proven, then it would imply the ground state of the Hubbard model with even numbers of electrons always has minimal pseudospin---a pseudospin tower.
These results are higher-dimensional extensions of the Lieb-Mattis theorem.

Significant work was also done on superconductivity. Kubo and Kishi showed that the spin susceptibility was bounded at all temperatures for the attractive Hubbard model, and the partial particle-hole transformation then showed that the on-site s-wave superconducting susceptibility and the charge-density-wave susceptibility are always bounded for the repulsive model. Note that it has nothing to say about whether a $d$-wave order might exist in the repulsive Hubbard model. This work then led to a proof of the superconducting off-diagonal long-range order in the attractive Hubbard model~\cite{tian2} and a discussion of so-called eta-pairing~\cite{shen3}. Tian also proved that the repulsive model had antiferromagnetic correlations~\cite{tian3}.

The examination of flat bands led to a flurry of activity on ferromagnetism led by Mielke~\cite{mielke1,mielke2} and Tasaki~\cite{tasaki1} and both~\cite{both}. Then, after graphene came onto the seen at the turn of the century, interest in flat bands and Lieb lattices really flourished. Some examples of work in this area include creating flat bands in optical lattices~\cite{optical}, in photonic lattices~\cite{photonics}, the relationship to superfluidity~\cite{superfluidity}, and the electronic Lieb-lattice has been created experimentally as well~\cite{expt}. The ferromagnetism~\cite{ferro} and topological properties~\cite{topo} have also been examined on real materials.

Finally, the theorems for the SU(2) Hubbard model, have been extended to the attractive SU(N) model~\cite{sun}.

The electron-phonon work has also spawned new theorems by others. Of note are the three papers by Miyao, that examine the Holstein model in one-dimension~\cite{miyao1}, the Su-Schrieffer-Heeger model in one dimension~\cite{miyao2}, and the long-range order in the Holstein model~\cite{miyao3}.

We end our discussion of the impact of these original ideas with a discussion of many-body localization. As Lieb showed in his original proof, there are two lattices of note---the original lattice $\Lambda$, which governs the single-particle hopping matrix, and the many-body lattice $\tilde\Lambda$, which governs the many-body states. In particular, if the original lattice is connected, then the many-body lattice is connected as well. Just like the ideas of localization in a single-particle problem is described by Anderson localization, localization on the many-body lattice describes the phenomena of many-body localization. While localization ideas were not discussed in Lieb's work, the idea to look into the properties of the many-body lattice become instrumental when analyzing many-body localization. For example, this is the analysis pursued by Roy and Logan~\cite{logan} for the many-body localization transition.

Hal Tasaki has written a book~\cite{tasaki-book} with three chapters covering material discussed here as well.

It is clear the work on spin-reflection positivity, and the other ideas presented in these works by Elliott Lieb are likely to have significant impact in the future as well.

\section{Historical context}

I will take a moment to describe the history of my interactions with Elliott Lieb. In 1984, I was a senior at Princeton University and I took a differential geometry class with Elliott in the math department. Then, when I was a graduate student in Physics at Berkeley, I made a transition from string theory to condensed matter physics. I read many of Elliott's works from the 1960's as I was preparing for work in this new field (for me). My first research project was on the Falicov-Kimball model, where we found a wide range of interesting ordered phases as functions of the electron density and the interaction strength. I would visit Princeton annually for reunions, and it was then that I started meeting Elliott whenever I would be on campus. The summer I was transitioning from a postdoctoral fellowship to a tenure-track job at Georgetown, I met with Elliott and discussed with him about the work I had been doing on the Holstein model in infinite dimensions. The next day, Elliott had written on the board the spin-reflection positivity argument for the existence of a spin-singlet state. The paper was finished shortly thereafter (with significant assistance by Jan-Philip Solovej). I spent a sabbatical visiting Princeton once a week where we worked with Daniel Ueltschi on proving phase separation in the Falicov-Kimball model---this was a result I had conjectured for many years, but was unable to make much progress on, until Elliott came up with the idea for how we define the boundary in the problem---from that point on, all became clear as to how we prove the phase separation. I have not worked with Elliott again for the past twenty years, but we keep in touch, and I always feel his influence in much of the work that I do.

\section{Conclusion}

What lies in the future? This is always hard to tell, but it is clear that we have not seen the last of the results related to spin-reflection positivity. The idea was a major breakthrough that has allowed a number of important results about strongly correlated models in condensed matter physics be proven. Such results always stand the test of time. But, it is highly likely that more remain on the horizon. We just need a little bit of inspiration, and a lot of perspiration, to be able to prove them!

%------
% Insert acknowledgments and information
% regarding funding at the end of the last
% section, i.e., right before the bibliography.
%------

\begin{ack}
This work would not have been possible had I not had the good fortune of collaborating with almost Elliott 30 years ago. It was a great experience and he was a fantastic collaborator. I first met Elliott as my instructor in a Differential Geometry course in 1984; he has been a great friend and source of inspiration ever since. 
\end{ack}

\begin{funding}
This work was supported by the Department of Energy, Office of Basic Energy Sciences, Division of Materials Sciences
and Engineering under Contract No. DE-FG02-08ER46542. It was also supported by the McDevitt bequest at Georgetown.
\end{funding}

%------
% Insert the bibliography.
%------

\end{document}